\documentclass[superscriptaddress,prl,twocolumn,nofootinbib]{revtex4}
\usepackage{amsmath,amssymb}
\usepackage{graphicx,epsfig}
\def\simgt{\mathrel{\lower2.5pt\vbox{\lineskip=0pt\baselineskip=0pt
           \hbox{$>$}\hbox{$\sim$}}}}
\def\simlt{\mathrel{\lower2.5pt\vbox{\lineskip=0pt\baselineskip=0pt
           \hbox{$<$}\hbox{$\sim$}}}}

\def\stacksymbols #1#2#3#4{\def\theguybelow{#2}
    \def\vp{\lower#3pt}
    \def\sp{\baselineskip0pt\lineskip#4pt}
    \mathrel{\mathpalette\intermediary#1}}

\def\intermediary#1#2{\vp\vbox{\sp
     \everycr={}\tabskip0pt
     \halign{$\mathsurround0pt#1\hfil##\hfil$\crcr#2\crcr
              \theguybelow\crcr}}}

\def\beq{\begin{equation}}
\def\eeq#1{\label{#1}\end{equation}}
\def\eeqn{\end{equation}}
\def\beqa{\begin{eqnarray}}
\def\eeqa#1{\label{#1}\end{eqnarray}}
\def\eeqan{\end{eqnarray}}

\begin{document}

\title{Hidden Gauge Symmetries: A New Possibility at the Colliders}

\author{Tianjun Li}
\affiliation{School of Natural Sciences, Institute for Advanced Study,
  Einstein Drive, Princeton, NJ 08540, USA}

\author{S. Nandi} 
\affiliation{Department of Physics, Oklahoma State University, 
Stillwater, OK 74078-3072, USA;\\ Fermi National Accelerator Laboratory,
P.O. Box 500, Batavia, IL 60510, USA}\thanks{S. Nandi was a Summer Visitor at Fermilab.}



\begin{abstract}

We consider a new physics possibility at the colliders: the
observation of TeV scale massive vector bosons 
in the non-adjoint representations under the Standard Model 
(SM) gauge symmetry. To have
a unitary and renormalizable theory, 
we propose a class of models with gauge symmetry
$\prod_i G_i \times SU(3)'_C \times SU(2)'_L \times U(1)'_Y$ where the
SM fermions and Higgs fields are singlets
under the hidden gauge symmetry $\prod_i G_i$,
and such massive vector bosons appear after the
gauge symmetry is spontaneously broken down to the SM gauge symmetry.
We discuss the model with $SU(5)$ hidden gauge symmetry in detail, and 
comment on the generic phenomenological implications.
\\ \\
{PACS numbers:\,14.65.Ha,\,12.60.-i\, \hfill [OSU-HEP-04-09, FERMILAB-PUB-04-146-T] }

\end{abstract}
\maketitle

{\em Introduction --}  The Standard Model (SM), based on the local
gauge symmetry $SU(3)_C\times SU(2)_L\times U(1)_Y$, is very
successful in describing all the experimental results below
the  TeV scale as an excellent effective field theory, although
it is widely believed that it is not the whole story.  
 Discovery of new particles is  highly anticipated at 
the colliders, especially
the Large Hadron Collider (LHC).
The most likely and reasonably
well motivated candidates are the Higgs bosons, superpartners of
ordinary particles, and the extra $Z'$ boson. However, 
it is important to think of other alternatives or entirely new 
possibilities before the LHC turns on.

In the SM, we have fermions (spin 1/2) and scalars 
(Higgs fields) (spin 0) which do not belong to adjoint 
representations under the SM gauge symmetry. 
Can we also have TeV scale  vector bosons (spin 1) which belong
to the non-adjoint representations under the SM 
gauge symmetry? Can we construct a renomalizable theory realizing
such a possibility? These are very interseting theoretical questions
that we will address in this work.
Discovery of these massive vector bosons
at the LHC will open up a new window for our understanding of
the fundamental theory describing the nature.

How can we construct a consistent  theory 
involving the massive vector bosons which do not belong to the adjoint
representations under the SM gauge symmetry?
If the massive vector bosons are not the gauge bosons of a symmetry group, 
there are some theoretical problems
from the consistency of quantum field theory, for instance, the
unitarity and renormalizability~\cite{Lee:1962vm}.
On the other hand,
 the gauge bosons are massless if the gauge symmetry
is exact. When the gauge symmetry is spontaneously
broken via the Higgs mechanism, the interactions of
the massive gauge bosons satisfy both the unitarity and the
renormalizability of the theory~\cite{Hooft,Lee}.
Thus, the massive vector bosons must be the gauge bosons arising from
the spontaneously gauge symmetry breaking.

As we know, a lot of models with extra 
TeV scale gauge bosons have been proposed previously in the
literature. However, those massive gauge bosons  either  belong
to the adjoint representations or are singlets
under the SM gauge 
symmetry~\cite{Hill:1991at,Hill:1993hs,Dicus:1994sw,
Muller:1996dj,Malkawi:1996fs,Erler:2002pr,PSLR}. For example, in the
top color model~\cite{Hill:1991at,Hill:1993hs,Dicus:1994sw}, 
the colorons belong to the adjoint representation
of the $SU(3)_C$; in the top 
flavor model~\cite{Muller:1996dj,Malkawi:1996fs}, the extra $W$ and $Z$
bosons belong to the adjoint representation of the $SU(2)_L$, while
in the $U(1)'$ model~\cite{Erler:2002pr}, 
the new $Z'$ boson is a singlet under the SM gauge symmetry.
In the Grand Unified Theories such as $SU(5)$ and
 $SO(10)$~\cite{Georgi:1974sy,Fritzsch:1974nn},
there are such kind of massive gauge bosons, however, their
masses are around the unification scale $\sim 10^{16}$ GeV 
to satisfy the proton decay constraints. 
Thus, we have to propose a new class of models.

When we embed the SM gauge groups
into larger groups (not necessarily a semi-simple group),
in general, we may have the massive gauge bosons that do not
belong to the adjoint representation under the SM gauge 
symmetry. However,
there are stringent constraints on the TeV scale massive gauge 
bosons from various experiments, for instance, the
Tevatron and the LEP.  To be consistent with all the current data,
we consider  a gauge symmetry ${\cal G} \equiv 
\prod_i G_i \times SU(3)'_C \times SU(2)'_L \times U(1)'_Y$.
The quantum numbers of the SM fermions and Higgs fields under
the $SU(3)'_C \times SU(2)'_L \times U(1)'_Y$ gauge symmetry are
the same as those of them under the SM gauge 
symmetry $SU(3)_C \times SU(2)_L \times U(1)_Y$,
while they are all singlets under $\prod_i G_i$. Hence $\prod_i G_i$ is 
the hidden gauge symmetry. After the 
gauge symmetry ${\cal G}$ is spontaneously broken down to
the SM gauge symmetry at the TeV scale via Higgs mechanism,
some of the massive gauge bosons from the ${\cal G}$ breaking
do not belong to the adjoint representations under the SM
gauge symmetry. 
These TeV scale gauge bosons 
may be observable at the LHC, Tevatron Run 2, and future
International Linear Collider (ILC),
and may give us an indication 
of a  new hidden gauge symmetry of the nature beyond
the SM gauge symmetry. Interestingly, in the string model
buildings, there generically exists the additional gauge symmetry
which can be considered as our hidden gauge
 symmetry~\cite{Cvetic:2004ui}.

For simplicity, we consider one hidden gauge group, 
{\it i.e.}, ${\cal G} \equiv  G \times SU(3)'_C \times SU(2)'_L \times U(1)'_Y$,
because the discussions for the general models are quite similar.
There are many choices for 
$G$, for example, $G= SU(2)$, $SU(3)$, $SU(4)$, $G_2$, $F_4$, $Sp(4)$, 
$Sp(6)$, $SU(5)$, $SO(10)$, $E_6$, etc. 
After the gauge symmetry ${\cal G}$ is spontaneously broken down to the SM
gauge symmetry,
 the possible non-adjoint representations of the massive gauge bosons
under the SM gauge symmetry $SU(3)_C \times SU(2)_L\times U(1)_Y$
in these models are
\begin{eqnarray}
&& ({\bf {1}}, {\bf 1}, {\bf  Q_1}) \oplus
({\bf {1}}, {\bf 1}, {\bf  -Q_1})~,~
({\bf {1}}, {\bf 2}, {\bf Q_2}) \oplus
({\bf {1}}, {\bf 2}, {\bf -Q_2}) ~,~ \nonumber\\ &&
({\bf {3}}, {\bf 1}, {\bf Q_3}) \oplus
({\bf {\bar 3}}, {\bf 1}, {\bf -Q_3})~,~
({\bf {3}}, {\bf 1}, {\bf 0}) \oplus
({\bf {\bar 3}}, {\bf 1}, {\bf 0})~,~\nonumber\\ &&
({\bf {6}}, {\bf 2}, {\bf  Q_4}) \oplus
 ({\bf {\bar 6}}, {\bf 2}, {\bf  -Q_4})~,~
({\bf {6}}, {\bf 1}, {\bf  Q_5}) \oplus
({\bf {\bar 6}}, {\bf 1}, {\bf  -Q_5})~,~\nonumber\\ &&
 ({\bf {\bar 3}}, {\bf 3}, {\bf  Q_6}) 
\oplus ({\bf {3}}, {\bf { 3}}, {\bf  -Q_6}) ~,~
({\bf {3}}, {\bf 2}, {\bf  Q_7}) \oplus
 ({\bf {\bar 3}}, {\bf 2}, {\bf - Q_7})~,~\nonumber\\ &&
({\bf 1}, {\bf 3}, {\bf {Q_8}}) \oplus
({\bf 1}, {\bf {3}}, {\bf { -Q_8}})~,~
\label{Allmgb}
\end{eqnarray} 
where $Q_i\not=0$.
In our models, there generically exist the massive
gauge bosons which belong to the adjoint representations
under the SM gauge symmetry. Because we are not interested
in the adjoint massive gauge bosons that have been 
studied previously~\cite{Hill:1991at,Hill:1993hs,Dicus:1994sw,
Muller:1996dj,Malkawi:1996fs}, we emphasize that we do not
consider them in this paper.

To be concrete, we shall give the formalism realizing our idea
for a simple model with $G=SU(5)$.

{\em Formalism --}  
We consider a model with 
${\cal G} \equiv SU(5)\times SU(3)'_C \times SU(2)'_L \times U(1)'_Y$
gauge symmetry where $SU(5)$ is a hidden gauge symmetry.
We denote the gauge fields
for $SU(5)$ and $SU(3)'_C \times SU(2)'_L \times U(1)'_Y $
as ${\widehat A}_{\mu}$ and ${\widetilde A}_{\mu}$, respectively,
and the gauge couplings for $SU(5)$, $SU(3)'_C$, $SU(2)'_L$
and $U(1)'_Y$ are $g_5$, $g'_3$, $g'_2$ and $g'_Y$,
respectively. The Lie algebra indices for the generators of
 $SU(3)$, $SU(2)$ and $U(1)$
are denoted by $a3$, $a2$ and $a1$, respectively, and
the  Lie algebra indices for the generators of
$SU(5)/(SU(3)\times SU(2)\times U(1))$ are denoted by ${\hat a}$.
After the $SU(5)\times SU(3)'_C \times SU(2)'_L \times U(1)'_Y $
gauge symmetry is broken down to the SM gauge symmetry
$SU(3)_C\times SU(2)_L \times U(1)_Y$,
 we denote the massless gauge fields for the
SM gauge symmetry as $A_{\mu}^{ai}$, and
the massive gauge fields as $B_{\mu}^{ai}$ and ${\widehat A}_{\mu}^{\hat a}$.
The gauge couplings for the SM gauge symmetry
$SU(3)_C$, $SU(2)_L$ and $U(1)_Y$ are $g_3$, $g_2$ and $g_Y$, respectively.

To break the $SU(5)\times SU(3)'_C \times SU(2)'_L \times U(1)'_Y $
gauge symmetry down to the SM gauge symmetry, we introduce
four bifundamental Higgs fields $U_1$, $U_2$, $U_3$ and $U_4$ with the 
following quantum numbers:

\begin{equation}
\label{SU(5)}
\begin{array}{|c|c|c|}
\hline
 {\rm Fields}     & SU(5) & SU(3)'_C \times SU(2)'_L \times U(1)'_Y \\
\hline
{\vrule height 15pt depth 5pt width 0pt}
 U_1       & {\bf 5} & ({\bf {\bar 3}}, {\bf 1}, {\bf {1/3}})\\
\hline
 U_2       & {\bf {\bar 5}} & ({\bf {3}}, {\bf 1}, {\bf -1/3})\\
\hline
 U_3       & {\bf 5} & ({\bf {1}}, {\bf { 2}}, {\bf -1/2})\\
\hline
 U_4       & {\bf {\bar 5}} & ({\bf {1}}, {\bf 2}, {\bf 1/2})\\
\hline
\end{array} \nonumber
\end{equation}
In order to break the $SU(5)\times SU(3)'_C \times SU(2)'_L \times U(1)'_Y $
gauge symmetry down to the SM gauge symmetry
$SU(3)_C\times SU(2)_L \times U(1)_Y$, we only need two
Higgs fields, 
one from the $U_1$ and $U_2$, and the other one from the $U_3$ and $U_4$.
The reason, for choosing four Higgs fields 
$U_1$, $U_2$, $U_3$ and $U_4$, is that
we can generalize this non-supersymmetric model to the supersymmetric model
without adding the new particle contents except the superpartners of
the particles.

To give the vacuum expectation values (VEVs)
 to the bifundamental Higgs fields $U_i$, we consider the
following potential
\begin{eqnarray}
V &=& \sum_{i=1}^4 
\left[ \lambda_i \left( |U_i|^2 -{\tilde v}_i^{ 2} \right)^2
+{\tilde m}_i^2 |U_i|^2 \right]
+\lambda_{12} | U_1 U_2 - {\tilde v}_{12}^{ 2} |^2
\nonumber\\ &&
+\lambda_{34} | U_3 U_4 - {\tilde v}_{34}^{ 2} |^2
+ \left[ {\tilde m}_{12}^2 U_1 U_2
+ {\tilde m}_{34}^2 U_3 U_4
\right.\nonumber\\ &&\left.
+ \lambda_{1234}  U_1 U_2 U_3 U_4
+ {{y_{13}}\over {M_{13}}} U_1 U_1 U_1 U_3 U_3
\right.\nonumber\\ &&\left.
+ {{y_{24}}\over {M_{24}}} U_2 U_2 U_2 U_4 U_4 
+ {\rm H.C.}\right]~,~\,
\label{potential}
\end{eqnarray} 
where $M_{13}$ and $M_{24}$ need not be at the GUT scale or Planck
scale, and can be the mass scales of the heavy
fields because the non-renormalizable operators can be obtained 
from the renormalizable operators by
integrating out the heavy fields. In addition,
there is no global symmetry in above potential
except the gauge symmetry 
$SU(5)\times SU(3)'_C \times SU(2)'_L \times U(1)'_Y $,
so, there is no Goldstone boson. 

We choose the following VEVs for the fields $U_i$
\begin{eqnarray}
 <U_1> =  {v_1} \left(
  \begin{array}{c}
    I_{3\times3} \\
    0_{2\times3} \\
  \end{array}
  \right)~, \quad
 <U_1> =  {v_2} \left(
  \begin{array}{c}
    I_{3\times3} \\
    0_{2\times3} \\
  \end{array}
  \right)~,
\end{eqnarray}
\begin{eqnarray}
 <U_3> =  {v_3} \left(
  \begin{array}{c}
    0_{3\times2} \\
    I_{2\times2} \\
  \end{array}
  \right)~, \quad
 <U_4> =  {v_4} \left(
  \begin{array}{c}
    0_{3\times2} \\
    I_{2\times2} \\
  \end{array}
  \right)~,
\end{eqnarray}
where $I_{i\times i}$ is the $i\times i$ identity matrix, and
$0_{i\times j}$ is the $i\times j$ matrix where all the entries are
zero. We asuume that $v_i$ ($i=1, 2, 3, 4$) are in the 
TeV range so that the massive gauge bosons  have TeV scale masses.

From the kinetic terms for the fields
$U_i$ , we obtain the mass terms for the gauge fields 
\begin{eqnarray}
&&\sum_{i=1}^4 \langle (D_{\mu} U_i)^{\dagger} D^{\mu} U_i \rangle
= {1\over 2} \left(v_1^2 +v_2^2 \right) \left( g_5 {\widehat A}_{\mu}^{a3} - 
g'_3 {\widetilde A}_{\mu}^{a3} \right)^2
\nonumber\\ &&
+ {1\over 2} \left(v_3^2 +v_4^2 \right) \left( g_5 {\widehat A}_{\mu}^{a2} - 
g'_2 {\widetilde A}_{\mu}^{a2} \right)^2
\nonumber\\ &&
+\left( {{v_1^2}\over 3} + {{v_2^2}\over 3} 
+ {{v_3^2}\over 2} + {{v_4^2}\over 2} \right) 
\left( g_5^Y {\widehat A}_{\mu}^{a1} - 
g'_Y {\widetilde A}_{\mu}^{a1} \right)^2
\nonumber\\ &&
+ {1\over 2} g_5^2 \left(v_1^2 +v_2^2 +v_3^2 +v_4^2 \right) 
\left(X_{\mu} \overline{X}_{\mu}
+ Y_{\mu} \overline{Y}_{\mu}\right)
~,~\,
\label{massterm}
\end{eqnarray}
 where $g_5^Y \equiv {\sqrt 3} g_5/{\sqrt 5}$,
 and we define the complex fields
($X_{\mu}$, $Y_{\mu}$) and (${\overline{X}_{\mu}}$, ${\overline{Y}_{\mu}}$)
with quantum numbers (${\bf 3}$, ${\bf 2}$, ${\bf -{5/6}}$) 
and (${\bf {\bar 3}}$, ${\bf 2}$, ${\bf {5/6}}$), respectively
from the gauge fields ${\widehat A}_{\mu}^{\hat a}$, similar to that
in the usual $SU(5)$ model~\cite{Georgi:1974sy}.

From the original gauge fields ${\widehat A}_{\mu}^{ai}$ and
${\widetilde A}_{\mu}^{ai}$ and from
Eq. (\ref{massterm}), we obtain the massless gauge bosons $A_{\mu}^{ai}$ and the
TeV scale massive gauge bosons $B_{\mu}^{ai}$ ($i=3, 2, 1$)
which are in the adjoint represenations of the SM gauge
symmetry
\begin{eqnarray}
\left(
\begin{array}{c}
A_\mu^{ai} \\
B_\mu^{ai}
\end{array} \right)=
\left(
\begin{array}{cc}
\cos\theta_i & \sin\theta_i \\
-\sin\theta_i & \cos\theta_i
\end{array}
\right)
\left(
\begin{array}{c}
{\widehat A}_\mu^{ai} \\
{\widetilde A}_\mu^{ai}
\end{array} \right)
~,~\,
\end{eqnarray}
where $i=3, 2, 1$, and 
\begin{eqnarray}
\sin\theta_j \equiv {{g_5}\over\displaystyle {\sqrt {g_5^2 +(g'_j)^2}}}
~,~
\sin\theta_1 \equiv {{g^Y_5}\over\displaystyle 
{\sqrt {(g_5^{Y})^2 +(g_Y^{\prime})^2}}} ~,~\,
\end{eqnarray}
where $j=3, 2$.
We also have the massive gauge bosons
($X_{\mu}$, $Y_{\mu}$) and (${\overline{X}_{\mu}}$, ${\overline{Y}_{\mu}}$)
which are not in the adjoint representations of the SM gauge symmetry.
 So, the $SU(5)\times SU(3)'_C \times SU(2)'_L \times U(1)'_Y $
gauge symmetry is  broken down to the diagonal SM gauge symmetry
$SU(3)_C\times SU(2)_L \times U(1)_Y$, and the theory is
unitary and renormalizable. The SM gauge couplings
$g_j$ ($j=3, 2$) and $g_Y$ are given by
\begin{eqnarray}
{1\over {g_j^2}} ~=~ {1\over {g_5^2}} + {1\over {(g'_j)^2}}~,~
{1\over {g_Y^2}} ~=~ {1\over {(g^Y_5)^2}} + {1\over {(g'_Y)^2}}~.~\,
\end{eqnarray}

If the theory is perturbative,  the upper and low
bounds on the gauge couplings $g_5$, $g'_3$, $g'_2$ and $g'_Y$ are 
\begin{eqnarray}
&& g_3 ~< ~g_5 ~< ~{\sqrt {4\pi}} ~,~ g_3 ~<~ g'_3 ~<~ {\sqrt {4\pi}} ~,~
\\ &&
g_2 ~<~ g'_2 ~<~
{{g_3 g_2}\over {\sqrt {g_3^2 -g_2^2}}}~,~ \\ &&
g_Y ~<~ g'_Y  ~<~
{{{\sqrt 3} g_3 g_Y}\over {\sqrt {3g_3^2 -5g_Y^2}}}~.~ \,
\end{eqnarray}
Note that the gauge coupling  $g_5$ for $SU(5)$  is naturally 
large at the TeV scale because
the beta function of $SU(5)$ is negative, {\it i.e.},
$SU(5)$ is asymptotically free.

{\em Phenomenological Implications --}
 The interactions among the gauge bosons of
$SU(5)\times SU(3)'_C \times SU(2)'_L \times U(1)'_Y $
can be obtained from the kinetic terms of the gauge fields.
For instance, the interactions between the SM gauge bosons
$A_{\mu}^{aj}$ and the massive gauge bosons $B_{\mu}^{aj}$
are given by
\begin{eqnarray}
-{\cal L}_{\rm gauge} & = & {1\over 2} g_j\left[ A^3+3AB^2+2\cot 2\theta
B^3\right] 
\nonumber\\ &&
+ {1\over 4} g_j^2 \left [A^4+6A^2B^2+4\left( 2\cot 2\theta
\right) AB^3 
\right.\nonumber\\ &&\left.
 +\left( \tan ^2\theta +\cot ^2\theta -1\right) B^4 \right]~,~\,
\label{L1}
\end{eqnarray}
where $j=2,3$.  The 
$A^3$ and $A^4$ represent schematically the usual three
and four point gauge interactions, respectively
\begin{eqnarray}
&& A^3\equiv f_{abc}
\left(\partial _\mu A_{\nu a}-\partial _\nu A_{\mu a} \right) A^{\mu}_b
A^{\nu}_c ~,~ \\ &&
\label{six}A^4 \equiv f_{abc}f_{ade}A_{\mu b}A_{\nu c}A_d^\mu A_e^\nu 
~.~ \,
\end{eqnarray}
And we choose the similar convention for the other terms. 
Note that a single heavy boson $B_{\mu}^{a3}$ does not 
 couple to two or three gluons, and hence can only be pair produced
via the gluon-gluon fusions, or $s$-chanel gluon exchanges,
or $t$-chanel $B_{\mu}^{a3}$ exchanges
 at the hadronic colliders such as the LHC.

Similarly, the gauge bosons $X^{\mu}$ and $Y^{\mu}$
will couple to the SM gauge bosons via the gauge kinetic terms
for the $SU(5)$ gauge bosons. So, they can be pair produced 
from the fusions of the SM gauge bosons, or $s$-chanel SM gauge boson exchanges,
or $t$-chanel $X^{\mu}$ and $Y^{\mu}$ exchanges in the colliders, for instance, the
gluon fusions for the hadronic colliders.

 The interactions between the gauge bosons and the
SM fermions can be obtained from the kinetic terms of the
SM fermions. 
For example, let us consider the quark doublet $Q_i$ with
quantum number (${\bf 1}$, ${\bf 3}$, ${\bf 2}$, ${\bf 1/6}$ ) under
the gauge symmetry $SU(5)\times SU(3)'_C \times SU(2)'_L \times U(1)'_Y $.
The Lagrangian  for $Q_i$ is
\begin{eqnarray}
-{\cal L} &=& {\overline{Q_i}} \gamma^{\mu} D_{\mu} Q_i~,~ \,
\end{eqnarray}
where
\begin{eqnarray}
D_{\mu} & \equiv & \partial_{\mu} -i g'_3 T^{a3} {\widetilde A}_{\mu}^{a3}
-i g'_2 T^{a2} {\widetilde A}_{\mu}^{a2}
-i {1\over 6} g'_Y  {\widetilde A}_{\mu}^{a1}
\nonumber\\ 
&=&
\partial_{\mu} 
-i g_3 T^{a3} \left( A_{\mu}^{a3} + \cot\theta_3 B_{\mu}^{a3} \right)
\nonumber\\ &&
-i g_2 T^{a2} \left( A_{\mu}^{a2} + \cot\theta_2 B_{\mu}^{a2} \right)
\nonumber\\ &&
-i {1\over 6} g_Y \left( A_{\mu}^{a1} + \cot\theta_1 B_{\mu}^{a1} \right)
~.~ \,
\end{eqnarray}

We emphasize that although the gauge symmetry 
$SU(5)\times SU(3)'_C \times SU(2)'_L \times U(1)'_Y $ is broken down 
to the SM gauge symmetry at TeV scale, there is no proton
decay problem in our model. The reason is that
the gauge bosons ($X_{\mu}$, $Y_{\mu}$) 
and (${\overline{X}_{\mu}}$, ${\overline{Y}_{\mu}}$)
can not couple to the SM fermions directly,
hence they will not produce the observable proton decay.
Moreover, note that the bifundamental Higgs fields $U_i$
are in the fundamental representation of $SU(5)$ while
the SM fermions are $SU(5)$ singlets,
hence  any operator,
which involves the SM fermions and the $U_i$ fields,
has at least two fermions and two $U_i$ fields.
Thus, any such operator has dimension 5 or higher,
and then will be suppressed by the cut-off scale in
the theory, which, for example, is the Planck scale.
Thus, the bifundamental Higgs fields $U_i$ will not
generate the proton decay problem.

The model proposed here has two kinds of the
new vector bosons with masses at TeV scale:
one kind of gauge bosons belongs to the adjoint
representations of the SM gauge symmetry,
while the other kind of gauge bosons
($X_{\mu}$, $Y_{\mu}$) and (${\overline{X}_{\mu}}$, ${\overline{Y}_{\mu}}$)
has quantum numbers (${\bf 3}$, ${\bf 2}$, ${\bf -{5/6}}$) 
and (${\bf {\bar 3}}$, ${\bf 2}$, ${\bf {5/6}}$) under
the SM gauge symmetry. Both kinds of massive
gauge bosons have many phenomenological implications
which can be tested at the upcoming LHC. In the following,
we list some of these phenomenological implications:

(i) The massive QCD type gauge bosons $B_{\mu}^{a3}$ belonging to the
(${\bf { 8}}$, ${\bf 1}$, ${\bf 0}$) representation
of the SM gauge symmetry will couple to the gluons, and
will be dominantly pair produced via the gluon-gluon 
fusions, or $s$-chanel gluon exchanges,
or $t$-chanel $B_{\mu}^{a3}$ exchanges
 at the LHC. They will decay dominantly
to $q \bar q$ pairs. Thus, the 4 jet high $p_T$ signal
will be significantly enhanced compaired to that in the SM
due to these heavy gluon productions.

(ii) The heavy electroweak type gauge bosons, 
with quantum numbers (${\bf { 1}}$, ${\bf 3}$, ${\bf 0}$)
and (${\bf { 1}}$, ${\bf 1}$, ${\bf 0}$) under
the SM gauge symmetry, will be singly produced at the LHC
via $q \bar q$ annihilations, as well as will be pair 
produced via the $WW$ and $ZZ$ fusions. The decays of the
heavy photon and $Z$ boson to the lepton pair
${ l}^+ { l}^-$ will
give  very clean signals.

(iii) The signals of these gauge bosons $B_{\mu}^{ai}$ ($i=3, 2, 1$)
are very different from those in the topcolor and topflavor
models, as our gauge bosons couple to all three families of
the SM fermions universally.

(iv) These TeV scale gauge bosons will produce the
off-shell effects in the future ILC, and will also
contribute to $g_{\mu} -2$.

(v) The gauge bosons in the non-adjoint representations
of the SM gauge symmetry
(such as $X_{\mu}$ and $Y_{\mu}$ type particles in this
simple model) can also be pair produced via the gluon fusions,
or $s$-chanel gluon exchanges,
or $t$-chanel massive gauge bosons
(for example, $X_{\mu}$ and $Y_{\mu}$) exchanges at the LHC. Because their
couplings to the SM fermions are highly suppressed and they can not
decay to the bifundamental Higgs fields $U_i$ due to the kinematics reason,
they will be meta-stable and behave like the stable heavy quarks
and anti-quarks. If produced at the LHC, they
can be detected by their ionizations as they pass 
through the detector.

(vi) Choosing the different groups for the hidden gauge symmetry $G$,
such as $SU(2)$, $SU(3)$ and $SU(4)$, etc, we can have 
the massive gauge bosons with different non-adjoint representations under the SM gauge
symmetry. Note that in some of these models, 
the massive gauge bosons and the bifundamental Higgs fields
do not cause the proton decay,
and thus the massive gauge bosons' couplings to the SM fermions via the bifundamental 
Higgs fields by the non-renormalizable operators need not
to be Planck scale suppressed. So, their productions
and subsequent decays might  give rise to the other interesting
signals at the LHC.

  Although we have discussed the phenomenological implications,
mostly for the LHC, our proposal has similar implications for the 
ongoing Tevatron Run 2, and off-shell effects in the future ILC. If
our proposed massive vector bosons are light enough, the color octet
($B_{\mu}^{a3}$) can be pair produced or the electroweak 
massive bosons ($B_{\mu}^{a1,2}$)
can be singly produced at the Tevatron Run 2, giving rise to similar signals
as at the LHC. Also the metastable $X_{\mu}$ and $Y_{\mu}$ bosons 
can be pair produced, and can be
searched by looking for their ionization tracks. We encourage the Run 2 Groups
to look for these signals.

{\em Conclusions --}
We have proposed the interesting new physics possibility at the colliders:
the observation of
TeV scale massive vector bosons belonging to the non-adjoint
representations under the
SM gauge symmetry. A class of the 
$\prod_i G_i \times SU(3)'_C \times SU(2)'_L \times U(1)'_Y$
models realizing this possibility is
presented, where such
vector bosons are generated when the 
$\prod_i G_i \times SU(3)'_C \times SU(2)'_L \times U(1)'_Y$
gauge symmetry is spontaneously broken down to the SM gauge symmetry via 
the Higgs mechanism. These theories are thus unitary as well as
renormalizable. The observation of such gauge bosons at the Tevatron Run 2,
the LHC, or the future ILC  will open up
a new window for our understanding of physics beyond the Standard Model.

{\em Acknowledgments --}
We would like to thank S. Adler and J. Lykken for helpful discussions.
SN wishes to thank the members of the School of Natural Sciences, Institute for
Advanced Study for their warm hospitality and support during his Summer Visit there
when this work was initiated. He also wishes to thank the members of the Theory 
Group at Fermilab for their warm hospitality and support during his Summer Visit
there when this work was completed.
The research of T. Li was supported by the
  National Science Foundation under Grant No.~PHY-0070928,
and the research of S. Nandi
 is supported in part by the Department of Energy 
Grants ~DE-FG02-04ER46140 and DE-FG02-04ER41306.



\begin{thebibliography}{99}

\bibitem{Lee:1962vm}
T.~D.~Lee and C.~N.~Yang,
Phys.\ Rev.\  {\bf 128}, 885 (1962).

\bibitem{Hooft}
G.~'t Hooft, Nucl.\ Phys.\ B {\bf 33}, 173 (1971);
G.~'t Hooft and M.~J.~G.~Veltman, 
Nucl.\ Phys.\ B {\bf 50}, 318 (1972).


\bibitem{Lee}
B.~W.~Lee and J.~Zinn-Justin,
Phys.\ Rev.\ D {\bf 5}, 3121 (1972);
Phys.\ Rev.\ D {\bf 5}, 3137 (1972);
Phys.\ Rev.\ D {\bf 5}, 3155 (1972);
Phys.\ Rev.\ D {\bf 7}, 1049 (1973).


\bibitem{Hill:1991at}
C.~T.~Hill,
Phys.\ Lett.\ B {\bf 266}, 419 (1991);
Phys.\ Lett.\ B {\bf 345}, 483 (1995).

\bibitem{Hill:1993hs}
C.~T.~Hill and S.~J.~Parke,
Phys.\ Rev.\ D {\bf 49}, 4454 (1994).

\bibitem{Dicus:1994sw}
D.~A.~Dicus, B.~Dutta and S.~Nandi,
Phys.\ Rev.\ D {\bf 51}, 6085 (1995).

\bibitem{Muller:1996dj}
D.~J.~Muller and S.~Nandi,
Phys.\ Lett.\ B {\bf 383}, 345 (1996).

\bibitem{Malkawi:1996fs}
E.~Malkawi, T.~Tait and C.~P.~Yuan,
Phys.\ Lett.\ B {\bf 385}, 304 (1996).


\bibitem{Erler:2002pr}
J.~Erler, P.~Langacker and T.~Li,
Phys.\ Rev.\ D {\bf 66}, 015002 (2002), and references therein.

\bibitem{PSLR}
J.~C.~Pati and A.~Salam,
Phys.\ Rev.\ D {\bf 10}, 275 (1974);
R.~N.~Mohapatra and J.~C.~Pati,
Phys.\ Rev.\ D {\bf 11}, 566 (1975);
G.~Senjanovic and R.~N.~Mohapatra,
Phys.\ Rev.\ D {\bf 12}, 1502 (1975).

\bibitem{Georgi:1974sy}
H.~Georgi and S.~L.~Glashow,
Phys.\ Rev.\ Lett.\  {\bf 32}, 438 (1974).

\bibitem{Fritzsch:1974nn}
H.~Georgi, {\it Particles and Fields}, 1974 (APS/DPF Williamsburg), 
ed. C. E. Carlson (AIP, New York, 1975) p.575;
H.~Fritzsch and P.~Minkowski,
Annals Phys.\  {\bf 93}, 193 (1975).

\bibitem{Cvetic:2004ui}
For example, M.~Cvetic, T.~Li and T.~Liu,
Nucl.\ Phys.\ B {\bf 698}, 163 (2004), and references therein.



\end{thebibliography}
\end{document}